\documentclass[11pt]{article}

\pdfoutput=1

\usepackage[letterpaper,margin=1.25in]{geometry}
\usepackage[T1]{fontenc}
\usepackage{microtype}
\usepackage[style=authoryear,giveninits=true,uniquename=false,uniquelist=false,maxbibnames=99]{biblatex}
\usepackage[title]{appendix}
\usepackage[colorlinks=true,citecolor=blue,urlcolor=blue]{hyperref}
\usepackage{amsmath}
\usepackage{amsthm}
\usepackage{amssymb}
\usepackage{algorithm}
\usepackage{algorithmicx}
\usepackage{booktabs}
\usepackage{tikz}
\usepackage{authblk}

\renewbibmacro{in:}{}
\DeclareNameAlias{sortname}{family-given}
\DeclareNameAlias{default}{family-given}

\newcommand*{\email}[1]{\href{mailto:#1}{\nolinkurl{#1}}}
\newtheorem{theorem}{Theorem}
\newtheorem{lemma}{Lemma}
\newtheorem{proposition}{Proposition}
\newtheorem{definition}{Definition}
\DeclareMathOperator{\st}{s.t.}
\DeclareMathOperator{\argmin}{arg\,min}
\DeclareMathOperator{\h}{H}
\DeclareMathOperator{\e}{E}
\DeclareMathOperator{\var}{Var}

\numberwithin{equation}{section}

\newcommand\keywords[1]{%
\begin{NoHyper}
\renewcommand\thefootnote{}\footnote{\emph{Keywords:} #1}%
\addtocounter{footnote}{-1}%
\end{NoHyper}
}

\newcounter{algsubstate}

\newenvironment{algsubstates}{
	\setcounter{algsubstate}{0}%
	\renewcommand{\State}{%
		\refstepcounter{algsubstate}%
		\Statex{\footnotesize\arabic{ALG@line}.\arabic{algsubstate}:}\space
	}
}
{}

\title{Robust subset selection}
\author{Ryan Thompson\thanks{Email: \email{ryan.thompson@monash.edu}}}
\affil{Department of Econometrics and Business Statistics, Monash University}

\begin{document}

\maketitle

\begin{abstract}
The best subset selection (or ``best subsets'') estimator is a classic tool for sparse regression, and developments in mathematical optimization over the past decade have made it more computationally tractable than ever. Notwithstanding its desirable statistical properties, the best subsets estimator is susceptible to outliers and can break down in the presence of a single contaminated data point. To address this issue, a robust adaption of best subsets is proposed that is highly resistant to contamination in both the response and the predictors. The adapted estimator generalizes the notion of subset selection to both predictors and observations, thereby achieving robustness in addition to sparsity. This procedure, referred to as ``robust subset selection'' (or ``robust subsets''), is defined by a combinatorial optimization problem for which modern discrete optimization methods are applied. The robustness of the estimator in terms of the finite-sample breakdown point of its objective value is formally established. In support of this result, experiments on synthetic and real data are reported that demonstrate the superiority of robust subsets over best subsets in the presence of contamination. Importantly, robust subsets fares competitively across several metrics compared with popular robust adaptions of continuous shrinkage estimators.
\end{abstract}


\section{Introduction}

We study the canonical linear regression model $Y=X\beta+\varepsilon$ with response $Y=(y_1,\ldots,y_n)^T\in\mathbb{R}^n$, predictors $X=(x_1,\ldots,x_n)^T\in\mathbb{R}^{n\times p}$, regression coefficients $\beta=(\beta_1,\ldots,\beta_p)^T\in\mathbb{R}^p$, and noise $\varepsilon=(\varepsilon_1,\ldots,\varepsilon_n)^T\in\mathbb{R}^n$. It is assumed that the response is centered and that the predictors are standardized. In the low-dimensional regime, where the number of predictors $p$ is smaller than the number of observations $n$, it is straightforward to estimate $\beta$ using the least squares estimator. However, in numerous contemporary statistical applications, $p$ can be (much) greater than $n$, in which case the least squares estimator is no longer statistically meaningful. One way to navigate such situations is to assume that the underlying model is sparse, i.e., to assume only a small fraction of the available predictors are important for explaining the response. Even when $p<n$, estimators that induce sparsity are useful because they err on the side of simplicity and interpretability. The best subset selection (or ``best subsets'') estimator is one of the earliest estimators that operates in the spirit of this idea, solving the constrained least squares problem:
\begin{equation}
\label{eq:bss}
\begin{split}
\underset{\beta}{\min}\quad&\frac{1}{2}\sum_{i=1}^n(y_i-x_i^T\beta)^2 \\
\st\quad&\quad\|\beta\|_0\leq k,
\end{split}
\end{equation}
where $k$ is an integer such that $0\leq k\leq\min(n-1,p)$, and the $\ell_0$-norm $\left\|\beta\right\|_0:=\sum_{j=1}^p1(\beta_j\neq 0)$ is the number of nonzero elements in $\beta$. Best subset selection is a combinatorial problem due to the sparsity constraint on $\beta$. Unlike other well-known sparsity-inducing estimators such as the Lasso \parencite{Tibshirani1996}, the best subsets estimator (via its sparsity constraint) directly controls the number of predictors in the model.

The optimization problem \eqref{eq:bss} suggests that to solve for the best subsets estimator, one must conduct a combinatorial search for the subset of (at most) $k$ predictors that yields the best linear representation of the response. Although the resulting estimator has favorable statistical properties in terms of estimation, prediction, and selection \parencite{Bunea2007,Raskutti2011,Shen2013,Zhang2014}, actually solving the (nonconvex) combinatorial problem is no small feat. In fact, finding the best subset(s) is an NP-hard problem \parencite{Natarajan1995}, and popular implementations such as the \texttt{R} package \texttt{leaps} do not scale well beyond $p\approx30$. However, in recent work, \textcite{Bertsimas2016a} showed that the best subsets problem \eqref{eq:bss} can be formulated and solved (to global optimality) as a \emph{mixed-integer program}, a class of mathematical optimization problems that has undergone remarkable advancements. In the last 10 years, for instance, the commercial mixed-integer solver \texttt{Gurobi} has experienced a nearly 60-fold hardware-independent speedup \parencite{GurobiOptimization2020}. When used in conjunction with warm starts from a projected gradient descent method, \textcite{Bertsimas2016a} showed that their mixed-integer optimization approach for best subsets can be applied to problems with dimensions as large as $p\approx1000$. This development represents the first time that the best subsets estimator has been tractable for contemporary high-dimensional data after at least 50 years of literature and has paved the way for exciting new research \parencite{Bertsimas2016,Mazumder2017a,Kreber2019,Bertsimas2020,Bertsimas2020a,Hastie2020,Hazimeh2020,Mazumder2020,Takano2020,Kenney2021}.

Despite the impressive developments in computational tools for best subset selection, certain fundamental limitations in the estimator itself remain. Particularly relevant to real-world applications is the robustness of best subsets to contamination in the data or lack thereof. Similar to the nonsparse least squares estimator, best subsets is highly susceptible to contamination in both the response and the predictors. Specifically, in the \emph{casewise contamination framework}, where a portion of the rows of $Y$ and $X$ are outliers, a single contaminated data point can have an arbitrarily severe effect on best subsets. The absence of robustness to casewise contamination is an important practical limitation of best subsets and raises doubts about the appropriateness of the estimator for many applications of sparse regression involving outliers, e.g., earnings forecasting \parencite{Wang2007}, analytical chemistry \parencite{Smucler2017}, and biomarker discovery \parencite{CohenFreue2019}. Although robust adaptions of other sparse estimators such as the Lasso have been studied fairly intensively \parencite{Rosset2007,Wang2007,Lambert-Lacroix2011,Alfons2013,Nguyen2013,Wang2013,Lozano2016,Smucler2017,Chang2018,Yang2018,Amato2021}, a lack of similar research is available on the topic of best subsets due to computational considerations. The objective of this paper is to address this gap.

\subsection{Robust subset selection}

In view of the preceding discussion, we study a robust adaption of best subset selection: robust subset selection (or ``robust subsets''). Motivated by ideas related to robust statistics and advances in mathematical optimization, robust subsets generalizes the problem of selection to include both predictors \emph{and} observations, leading to the combinatorial problem:
\begin{equation}
\label{eq:rss}
\begin{split}
\underset{\beta,I}{\min}\quad&\frac{1}{2}\sum_{i\in I}(y_i-x_i^T\beta)^2 \\
\st\quad&\|\beta\|_0\leq k \\
&I\subseteq[n] \\
&|I|\geq h,
\end{split}
\end{equation}
where $h$ is an integer such that $k\leq h\leq n$, and the notation $[n]$ denotes the set $\{1,\ldots,n\}$. In effect, robust subsets performs a best subsets fit on the $h$ observations that produce the smallest square error while the most anomalous $n-h$ observations are ``trimmed.'' The idea of trimming anomalous observations is inspired by the method of \emph{least trimmed squares} (LTS), an estimator that is highly resistant to contamination in both $Y$ and $X$ and is well-established in the robust statistics literature \parencite{Rousseeuw1984}. Because minimizing the sum of squares in \eqref{eq:rss} without the sparsity constraint on $\beta$ leads to the LTS estimator, robust subsets can be interpreted as subset selection under LTS loss.

Although solving \eqref{eq:rss} exactly is theoretically intractable (it is NP-hard), this paper demonstrates that modern methods from mathematical optimization can be applied to tackle practical-sized problem instances with $n$ and $p$ in the hundreds, including the high-dimensional case when $p\gg n$. The resulting estimator is shown to have favorable statistical properties, both theoretically in terms of its finite-sample breakdown point and empirically in terms of its performance on synthetic and real data. Unlike robust adaptions of sparse estimators that rely on continuous shrinkage, the robust subsets estimator (via its nonconvex sparsity constraint on $\beta$) exhibits excellent support recovery and produces fitted models with few nonzeros.

\subsection{Contributions and organization}

The contributions of this work are summarized as follows. We first show that the problem of robust subset selection is amenable to formulation as a mixed-integer program, allowing us to leverage advancements in mixed-integer solvers to compute exact solutions. To complement this approach, we develop tailored heuristics to quickly obtain good feasible solutions to the robust subsets problem. These heuristics include a projected block-coordinate gradient descent method, for which we derive convergence properties, and a neighborhood search method, which exploits neighborhood information across a grid of the parameters $k$ and $h$ to generate an entire set of fitted models. Our heuristics can also rapidly generate warm start solutions that the solver can refine to produce high-quality fitted models. The breakdown point for the objective value of the robust subsets estimator is subsequently derived. Finally, numerical experiments are conducted on synthetic data under a comprehensive set of contamination settings. A real data application is also illustrated. Our implementation \texttt{robustsubsets} is made available as an open-source \texttt{R} package.

This paper is structured as follows. Section \ref{sec:background} provides a brief review of related work. Section \ref{sec:computational} introduces computational methods for robust subset selection. Section \ref{sec:breakdown} discusses robustness properties vis-\`{a}-vis the breakdown point. Section \ref{sec:experiments} presents results from numerical experiments. Section \ref{sec:archaeological} describes a real data application. Section \ref{sec:concluding} closes the paper with concluding remarks. Proofs are provided in the appendix.

\section{Background}
\label{sec:background}

In light of the extensive literature on the topics of sparse and robust regression (and their intersection), we provide a brief review of select work related to the present paper.

\subsection{Best subset selection and the Lasso}

Since at least the 1960s, best subset selection has been recognized as an important problem in statistics \parencite{Garside1965,Beale1967,Hocking1967}. \textcite{Furnival1974}, in a seminal paper, introduced an exact algorithm for best subsets that relies on a branch-and-bound method and is still used today in \texttt{leaps}.\footnote{See also \textcite{Gatu2006,Hofmann2007} for later, related work.} In more recent decades, computationally friendlier estimators have arisen, most notably the Lasso. Unlike best subsets, the Lasso is defined by a relatively simple convex problem:
\begin{equation}
\label{eq:lasso}
\begin{split}
\underset{\beta}{\min}\quad&\frac{1}{2}\sum_{i=1}^n(y_i-x_i^T\beta)^2 \\
\st\quad&\quad\|\beta\|_1\leq t,
\end{split}
\end{equation}
where $t>0$ controls the level of sparsity (albeit, indirectly). Any modern convex solver can optimize \eqref{eq:lasso} or its popular Lagrangian form:
\begin{equation}
\label{eq:lasso2}
\underset{\beta}{\min}\quad\frac{1}{2}\sum_{i=1}^n(y_i-x_i^T\beta)^2+\lambda\|\beta\|_1,
\end{equation}
which is equivalent to \eqref{eq:lasso} for some $\lambda>0$. In addition to convex solvers, efficient algorithms also exist for computing the Lasso that exploit its highly structured nature, including least angle regression \parencite[LARS,][]{Efron2004} and pathwise coordinate descent \parencite{Friedman2007}. Other noteworthy estimators include the Dantzig selector \parencite{Candes2007}, as well as those based on nonconvex penalties such as the smoothly clipped absolute deviation penalty \parencite{Fan2001} and the minimax concave penalty \parencite{Zhang2010}. For relevance, we limit our discussion to best subsets and the Lasso due to the prevalence of the latter in the robust statistics literature.

The Lasso problems \eqref{eq:lasso} and \eqref{eq:lasso2} are a convex relaxation of the best subsets problem \eqref{eq:bss}; they replace the $\ell_0$ constraint with a convex surrogate $\ell_1$ constraint (or penalty). Therefore, the Lasso is often interpreted as a heuristic for best subsets. However, unlike the best subsets parameter $k$, the Lasso parameters $t$ or $\lambda$ do not directly control the model sparsity. The Lasso also induces shrinkage on the regression coefficients, which can help or hinder depending on the level of noise \parencite{Hastie2020,Mazumder2020}. On the other hand, best subsets allows predictors to enter the model with a full least squares fit, removing the effect of other correlated predictors in the process.

From a theory standpoint, the Lasso requires that somewhat restrictive conditions hold to achieve good statistical properties \parencite[see, e.g.,][]{vandeGeer2009}. \textcite{Zhao2006} showed that the Lasso is only capable of selecting the true model consistently under the so-called irrepresentable condition, which places rather strong restrictions on the covariance of the predictors. \textcite{Zhang2014} derived bounds on prediction loss from the (thresholded) Lasso under a restricted eigenvalue condition on the predictor matrix. Even when this condition is satisfied, they showed that a substantial gap can still occur compared with the prediction loss from best subsets.

The empirical performance of best subsets compared with that of the Lasso is somewhat less well understood than the theory. \textcite{Bertsimas2016a} and \textcite{Hastie2020} performed empirical comparisons of the estimators. Perhaps unsurprisingly, neither estimator was found to dominate uniformly, but their experiments validated the stylistic fact that best subsets tends to produce fitted models that are significantly sparser than those from the Lasso, especially when the number of predictors is large. The simulations in this paper yield a similar finding in the contaminated setting in which the sparser models produced by robust subsets contain substantially fewer false positives than models generated by robust adaptions of the Lasso.

\subsection{Sparse and robust regression}

Like sparse regression, robust regression is a classical topic in statistics. A recent and detailed treatment of the subject is available in \textcite{Maronna2019}. With the proliferation of high-dimensional datasets, estimators that are simultaneously robust and sparse have become a topic of intense interest in recent years. In particular, a fairly extensive body of literature is available in robust statistics on the topic of the Lasso. One of the earliest papers in this area is \textcite{Wang2007}, which introduced the Lasso with least absolute deviation (LAD) loss:
\begin{equation}
\label{eq:ladlasso}
\underset{\beta}{\min}\quad\sum_{i=1}^n|y_i-x_i^T\beta|+\lambda\|\beta\|_1.
\end{equation}
Huber loss was used by \textcite{Rosset2007} and \textcite{Lambert-Lacroix2011}, and their estimators are related to the so-called extended Lasso \parencite{Nguyen2013}. Indeed, the Lasso has been studied under numerous other loss functions including exponential square loss \parencite{Wang2013}, minimum distance loss \parencite{Lozano2016}, and Tukey's bisquare loss \parencite{Smucler2017,Chang2018}. \textcite{Alfons2013} studied the Lasso with LTS loss, which effectively relaxes the nonconvex sparsity constraint on $\beta$ in the robust subsets problem \eqref{eq:rss} with an $\ell_1$ penalty. In a slightly different line of work, \textcite{Khan2007} robustified the LARS procedure by utilizing resistant estimators of mean and covariance. \textcite{Chen2013} proposed a related idea whereby the Lasso objective is restated in terms of trimmed inner products.

The body of literature studying best subsets in the contaminated setting is relatively limited. \textcite{Bertsimas2016a} showed that their optimization framework can incorporate subset selection with LAD loss, the $\ell_0$ constrained analogy of \eqref{eq:ladlasso}. However, unlike the robust subsets problem, the LAD subsets problem does not result in an estimator resistant to contamination in the predictor matrix, which is arguably the most relevant case for contemporary applications involving large numbers of predictors. Heuristic algorithms for trimmed $\ell_0$ constrained regression were developed and analyzed in \textcite{Bhatia2015} and \textcite{Suggala2019} under particular conditions on the predictor matrix. Unfortunately, the algorithmic frameworks developed in those papers are relevant only for contamination in the response and do not apply to the problems we consider wherein $X$ may also be contaminated. \textcite{Liu2020} studied related heuristics based on the interesting idea of using a robust estimate of the gradient. A drawback of this approach is that each algorithmic iteration involves solving an optimization problem afresh. After this work was first shared online, \textcite{Insolia2020} established some theoretical guarantees in the form of oracle properties for robust subset selection.

Robust regression is an inherently nonconvex problem; see, e.g., the discussion in \textcite{She2011}. \textcite{Alfons2013} showed that the use of a convex loss function with the Lasso, such as in the LAD Lasso problem, leads to a finite-sample breakdown point of $1/n$, the same as that from standard least squares loss. Accordingly, many sparse and robust estimators are defined in terms of nonconvex optimization problems, nearly all of which rely solely on heuristics that are only capable of delivering approximate solutions \parencite{Alfons2013,Smucler2017,Chang2018}. Although we also apply heuristics to a highly nonconvex problem, they form part of a broader framework that incorporates mixed-integer optimization which is guaranteed to converge to a global minimizer, if one exists. In the nonsparse setting, mixed-integer optimization has been used successfully in earlier works to find global minimizers for the problems of least trimmed squares \parencite{Zioutas2009} and its cousin least median of squares \parencite{Bertsimas2014}. See also \textcite{Hofmann2010} for a tailored branch-and-bound method for least trimmed squares that comes with global convergence guarantees.

\section{Computational methods}
\label{sec:computational}

This section details computational methods for robust subset selection. We begin with a brief primer on mixed-integer optimization and proceed to describe a mixed-integer program for robust subsets. Heuristics are presented, including a projected block-coordinate gradient descent method and a neighborhood search method. The heuristics, together with mixed-integer optimization, form a powerful computational toolkit for robust subsets. The section closes with practical guidance on parameter choices.

\subsection{Mixed-integer optimization}

\subsubsection{Primer}

Recall that the general form for a mixed-integer program with a quadratic objective, linear constraints, and variable $\omega\in\mathbb{R}^p$ is
\begin{equation}
\label{eq:mio}
\begin{split}
\underset{\omega}{\min}\quad&\omega^TQ\omega+q^T\omega \\
\st\quad&A\omega\leq b \\
&l_j\leq\omega_j\leq u_j,\quad j\in[p] \\
&\omega_j\in\mathbb{Z},\qquad\text{for some}~j\in[p], \\
\end{split}
\end{equation}
where positive semidefinite $Q\in\mathbb{R}^{p\times p}$ and $q\in\mathbb{R}^p$ form the objective, $A\in\mathbb{R}^{m\times p}$ is a constraint matrix and $b\in\mathbb{R}^m$ is a right-hand side vector, and $l\in\mathbb{R}^p$ and $u\in\mathbb{R}^p$ are lower and upper bound vectors. The program \eqref{eq:mio} is said to be a mixed-integer quadratic program, and the constraints $\omega_j\in\mathbb{Z}$ are said to be integrality constraints. It is these integrality constraints that render the feasible set nonconvex and lead to problems of the form \eqref{eq:mio} being NP-hard to solve in general \parencite{Aardal2005}. State-of-the-art mixed-integer solvers such as \texttt{CPLEX}, \texttt{GLPK}, \texttt{Gurobi}, and \texttt{MOSEK} optimize \eqref{eq:mio} by applying branch-and-bound methods in combination with cutting plane generation techniques and elaborate heuristics. Roughly speaking, branch-and-bound methods operate by reducing the original problem to a series of subproblems represented as a search tree. Branches of the search tree are enumerated only if they can improve on the incumbent solution, as determined by the estimated lower and upper bounds on the optimal value of the objective function. Optimality of the solution is declared once the lower and upper bounds converge.

\subsubsection{Mixed-integer programs}

We turn our attention to formulating robust subset selection as a mixed-integer program. Towards this end, we begin with a formulation for best subset selection, itself a special case of robust subset selection. Letting $s\in\{0,1\}^p$ be an auxiliary binary variable, the best subsets problem \eqref{eq:bss} has the following mixed-integer program representation:
\begin{equation}
\label{eq:bssmio}
\begin{split}
\underset{\beta,s}{\min}\quad&\frac{1}{2}\sum_{i=1}^n(y_i-x_i^T\beta)^2 \\
\st\quad&s_j\in\{0,1\},\qquad j\in[p] \\
&-s_j\mathcal{M}_{\beta}\leq\beta_j\leq s_j\mathcal{M}_{\beta},\qquad j\in[p] \\
&\sum_{j=1}^ps_j\leq k,
\end{split}
\end{equation}
where $\mathcal{M}_\beta>0$ is a problem-specific (fixed) parameter. The formulation \eqref{eq:bssmio} exploits the ``Big-M'' constraints $-s_j\mathcal{M}_{\beta}\leq\beta_j\leq s_j\mathcal{M}_{\beta}$ to enforce sparsity on $\beta$. These Big-M constraints have the effect that
\begin{equation*}
s_j=0\implies\beta_j=0\quad\text{and}\quad s_j=1\implies\beta_j\in[-\mathcal{M}_\beta,\mathcal{M}_\beta].
\end{equation*}
Hence, via $\mathcal{M}_\beta$, the $s_j$ act as switches that control whether the $\beta_j$ can take on nonzero values. The constraint $\sum_{j=1}^ps_j\leq k$ has the effect of upper bounding the $\ell_0$-norm of $\beta$:
\begin{equation*}
\sum_{j=1}^ps_j\leq k\implies\|\beta\|_0\leq k,
\end{equation*}
thereby yielding the desired level of sparsity in the fitted model.

To generalize the program \eqref{eq:bssmio} to solve for the problem of interest, we exploit the following (equivalent) reformulation of the robust subsets problem \eqref{eq:rss}:
\begin{equation}
\label{eq:rss2}
\begin{split}
\underset{\beta,\eta}{\min}\quad&\frac{1}{2}\sum_{i=1}^n(y_i-x_i^T\beta-\eta_i)^2 \\
\st\quad&\|\beta\|_0\leq k \\
&\|\eta\|_0\leq n-h,
\end{split}
\end{equation}
where we optimize over the continuous variable $\eta\in\mathbb{R}^n$ in place of the set-valued variable $I$. The indices of the nonzero elements in $\eta$ correspond to the complement of the set of observation indices $I$. The trick of introducing auxiliary variables to achieve robustness has been used in several works previously \parencite{McCann2007,Menjoge2010,She2011,Nguyen2013,Suggala2019}. In particular, it is known that constraining the $\ell_1$-norm of $\eta$ is equivalent to using Huber loss \parencite{She2011}, whereas constraining its $\ell_0$-norm is equivalent to using LTS loss \parencite{Suggala2019}. Though this trick has been used before, the algorithmic framework we develop and apply it in is itself novel.

To represent \eqref{eq:rss2} as a mixed-integer program, we introduce the auxiliary binary variable $z\in\{0,1\}^n$ to construct the following formulation:
\begin{equation}
\label{eq:rssmio}
\begin{split}
\underset{\beta,\eta,s,z}{\min}\quad&\frac{1}{2}\sum_{i=1}^n(y_i-x_i^T\beta-\eta_i)^2 \\
\st\quad&s_j\in\{0,1\},\qquad j\in[p] \\
&-s_j\mathcal{M}_{\beta}\leq\beta_j\leq s_j\mathcal{M}_{\beta},\qquad j\in[p] \\
&\sum_{j=1}^ps_j\leq k \\
&z_i\in\{0,1\},\qquad i\in[n] \\
&-z_i\mathcal{M}_{\eta}\leq\eta_i\leq z_i\mathcal{M}_{\eta},\qquad i\in[n] \\
&\sum_{i=1}^nz_i\leq n-h,
\end{split}
\end{equation}
where $\mathcal{M}_\eta>0$ is a Big-M parameter for $\eta$. The robust subsets program \eqref{eq:rssmio} enforces sparsity on both $\beta$ and $\eta$ via the Big-M constraints.

For \eqref{eq:rssmio} to be a valid formulation of the robust subsets problem, insofar as its optimal solution is the same as that of \eqref{eq:rss2}, the Big-M parameters must be sufficiently large. More precisely, $\mathcal{M}_\beta$ and $\mathcal{M}_\eta$ should satisfy $\mathcal{M}_\beta\geq\|\beta^\star\|_\infty$ and $\mathcal{M}_\eta\geq\|\eta^\star\|_\infty$ for $\beta^\star$ and $\eta^\star$ that is an optimal solution to \eqref{eq:rss2}. Alternatively, the specification of either of these parameters can be avoided by replacing the Big-M constraints with indicator constraints or special ordered set (SOS) constraints. Such constraints do not require specification of any parameters but have the same effect as Big-M constraints. An SOS constraint (of type 1) has the effect that
\begin{equation*}
\quad (\beta_j,1-s_j):\text{SOS-1}\implies \beta_j(1-s_j)=0.
\end{equation*}
Thus, replacing the Big-M constraints in \eqref{eq:rssmio} with SOS constraints does not change the optimal solution. However, it is our experience that the performance of the solver is generally superior when the problem is formulated with Big-M constraints. At the conclusion of this section, we show how the heuristics presented next can be used to estimate $\mathcal{M}_\beta$ and $\mathcal{M}_\eta$.

We make two remarks pertaining to the computation of a solution to \eqref{eq:rssmio}:
\begin{itemize}
\item For a given problem instance, it might be the case that $\beta$ is presumed to be fully dense (i.e., $k=p$), and it is thus desirable to remove the variable $s$ and the corresponding Big-M constraints from the formulation. Likewise, if it is presumed that the data are uncontaminated (i.e., $h=n$), it is helpful to remove $z$ from the problem, as well as $\eta$. The presolve routines used in most modern solvers are capable of identifying these situations and simplifying the formulation.
\item The high-level solution strategy used by the solver can usually be tuned. Often, the competing goals of (a) finding a new feasible solution and (b) proving optimality of the incumbent solution are balanced. Still, work on either of these goals can also be prioritized. This capability is particularly useful for day-to-day data-analytic work in which obtaining high-quality solutions with low runtime is principally of interest.
\end{itemize}

Finally, several techniques are available that can improve the performance of the solver via simple modifications to the mixed-integer program \eqref{eq:rssmio}. A brief discussion of these techniques is included in the appendix.

\subsection{Heuristics}

Although modern mixed-integer solvers are capable of solving the mixed-integer programs that we have presented, they are not (in general) sufficiently quick to be of use by themselves for practical-sized problem instances (e.g., $n$ and $p$ in the hundreds). To this end, we propose tailored heuristic methods that complement the mixed-integer optimization approach in the following ways:
\begin{itemize}
\item They can rapidly generate good feasible solutions to the robust subsets problem \eqref{eq:rss2} that can be exploited by the solver as warm starts.
\item Their solutions can be used to derive suitable values of the Big-M parameters (e.g., $\mathcal{M}_\beta$) required in the mixed-integer program.
\item They can be applied to cross-validate the parameters $k$ and $h$ with low computational cost, which might otherwise require multiple expensive calls to the solver.
\end{itemize}

\subsubsection{Projected block-coordinate gradient descent}

In general, provably exact minimizers to the robust subsets problem \eqref{eq:rss2} are unattainable without mixed-integer optimization. However, first-order optimization algorithms, namely, projected gradient descent methods, have been applied with great success to find good local minimizers for best subsets and related problems \parencite{Bertsimas2016a,Kudo2020,Mazumder2020}. Motivated by this success, we extend the projected gradient descent method developed in \textcite{Bertsimas2016a} for the best subsets problem \eqref{eq:bss} to the robust subsets problem \eqref{eq:rss2}. Their method involves a standard gradient descent update to the full set of coordinates followed by projection onto the feasible set of $k$-sparse solutions. Because the robust subsets problem is characterized by distinct blocks of coordinates, we adapt this scheme to perform block-coordinate updates. The resulting projected \emph{block-coordinate} gradient descent method finds good feasible solutions that yield upper bounds to the optimal value of the robust subsets objective function.

For simplicity of exposition, $f(\beta,\eta)$ is used to denote the objective function in \eqref{eq:rss2}:
\begin{equation}
\label{eq:obj}
f(\beta,\eta):=\frac{1}{2}\sum_{i=1}^n(y_i-x_i^T\beta-\eta_i)^2=\frac{1}{2}\|Y-X\beta-\eta\|_2^2.
\end{equation}
The objective function \eqref{eq:obj} has the partial derivatives 
\begin{equation*}
\nabla_\beta f(\beta,\eta)=-X^T(Y-X\beta-\eta)
\end{equation*}
and
\begin{equation*}
\nabla_\eta f(\beta,\eta)=-(Y-X\beta-\eta).
\end{equation*}
Observe that $\nabla_\beta f(\beta,\eta)$ and $\nabla_\eta f(\beta,\eta)$ are Lipschitz continuous, i.e., there exist real constants $L_\beta>0$ and $L_\eta>0$ such that
\begin{equation}
\label{eq:lipschitz1}
\|\nabla_\beta f(\beta,\eta)-\nabla_\beta f(\tilde{\beta},\eta)\|_2\leq L_\beta\|\beta-\tilde{\beta}\|_2\quad\forall\,\beta,\tilde{\beta}\in\mathbb{R}^p,\,\eta\in\mathbb{R}^n
\end{equation}
and
\begin{equation}
\label{eq:lipschitz2}
\|\nabla_\eta f(\beta,\eta)-\nabla_\eta f(\beta,\tilde{\eta})\|_2\leq L_\eta\|\eta-\tilde{\eta}\|_2\quad\forall\,\eta,\tilde{\eta}\in\mathbb{R}^n,\,\beta\in\mathbb{R}^p.
\end{equation}
In particular, the Lipschitz constants $L_\beta=\|X^TX\|_2$ and $L_\eta=1$, where $\|\cdot\|_2$ denotes the spectral norm of the matrix. The Lipschitz continuity of $\nabla_\beta f(\beta,\eta)$ and $\nabla_\eta f(\beta,\eta)$ leads to the block descent lemma \parencite{Beck2015}, whereby \eqref{eq:obj} can be upper bounded as follows.
\begin{lemma}
\label{lemma:descent}
Let $f(\beta,\eta)$ be the robust subset selection objective function \eqref{eq:obj}. Then, for any $\bar{L}_\beta\geq L_\beta$ and any $\bar{L}_\eta\geq L_\eta$, it holds that
\begin{equation*}
f(\tilde{\beta},\eta)\leq Q(\tilde{\beta},\beta):=f(\beta,\eta)+\nabla_\beta f(\beta,\eta)^T(\tilde{\beta}-\beta)+\frac{1}{2}\bar{L}_\beta\|\tilde{\beta}-\beta\|_2^2\quad\forall\,\beta,\tilde{\beta}\in\mathbb{R}^p,\,\eta\in\mathbb{R}^n
\end{equation*}
and
\begin{equation*}
f(\beta,\tilde{\eta})\leq R(\tilde{\eta},\eta):=f(\beta,\eta)+\nabla_\eta f(\beta,\eta)^T(\tilde{\eta}-\eta)+\frac{1}{2}\bar{L}_\eta\|\tilde{\eta}-\eta\|_2^2\quad\forall\,\eta,\tilde{\eta}\in\mathbb{R}^n,\,\beta\in\mathbb{R}^p.
\end{equation*}
\end{lemma}
The proposed projected block-coordinate gradient descent method performs cyclic updates by alternating between minimization of the upper bounds $Q(\tilde{\beta},\beta)$ and $R(\tilde{\eta},\eta)$. The hard-thresholding operator is pivotal to this minimization and is defined for a vector $c\in\mathbb{R}^p$ as
\begin{equation*}
\h(c;k)\in\underset{\alpha\in\mathbb{R}^p:\|\alpha\|_0\leq k}{\argmin}~\|\alpha-c\|_2^2.
\end{equation*}
Taking $\{(1),\ldots,(p)\}$ to denote an ordering of $\{1,\ldots,p\}$ such that $|c_{(1)}|\geq|c_{(2)}|\geq\cdots\geq|c_{(p)}|$, it is well-known that $\h(c;k)$ has the following analytic form:
\begin{equation*}
\hat{\alpha}_j=
\begin{cases}
c_j & \text{if}~j\in\{(1),\ldots,(k)\} \\
0 & \text{otherwise}
\end{cases},\qquad j\in[p].
\end{equation*}
The operator $\h(c,k)$ retains the $k$ largest elements of the vector $c$ measured in absolute value and sets the remaining elements to zero. Observe that $\h(c,k)$ is a set-valued map because more than one valid permutation of the indices might exist. Using the hard-thresholding operator, the computation for a single update to $\beta$ can be written as
\begin{equation*}
\begin{split}
\hat{\beta}&\in\underset{\tilde{\beta}\in\mathbb{R}^p:\|\tilde{\beta}\|_0\leq k}{\argmin}~Q(\tilde{\beta},\beta) \\
&=\underset{\tilde{\beta}\in\mathbb{R}^p:\|\tilde{\beta}\|_0\leq k}{\argmin}~\left\|\tilde{\beta}-\left(\beta-\frac{1}{\bar{L}_\beta}\nabla_\beta f(\beta,\eta)\right)\right\|_2^2 \\
&=\h\left(\beta-\frac{1}{\bar{L}_\beta}\nabla_\beta f(\beta,\eta);k\right).
\end{split}
\end{equation*}
Thus, with fixed $\eta$, an update to $\beta$ is performed by taking a gradient descent step followed by a mapping to the nearest $k$-sparse subspace of $\mathbb{R}^p$. The second set of coordinates $\eta$ can be updated similarly. In fact, with fixed $\beta$, such an update yields exact minimization with respect to $\eta$. This result follows from the definition of the hard-thresholding operator (take $\alpha=\eta$ and $c=Y-X\beta$).

Using the above ingredients, Algorithm \ref{alg:pbgd} presents the projected block-coordinate gradient descent method for optimization of \eqref{eq:rss2}.
\begin{algorithm}[H]
\caption{Projected block-coordinate gradient descent}
\label{alg:pbgd}
\hspace*{\algorithmicindent} \textbf{Input:} $\bar{L}_\beta\geq L_\beta$, $\bar{L}_\eta\geq L_\eta$, and $\epsilon>0$. \\
\hspace*{\algorithmicindent} \textbf{Initialize:} $\beta^{(0)}\in\mathbb{R}^k\times\{0\}^{p-k}$ and $\eta^{(0)}\in\mathbb{R}^{n-h}\times\{0\}^h$.
\begin{algorithmic}[1]
\State For $m\geq 0$, repeat the following steps until $f(\beta^{(m)},\eta^{(m)})-f(\beta^{(m+1)},\eta^{(m+1)})\leq\epsilon$:
\begin{algsubstates}
\State Update $\beta^{(m)}$ as
\begin{equation*}
\beta^{(m+1)}\in\h\left(\beta^{(m)}-\frac{1}{\bar{L}_{\beta}}\nabla_{\beta}f(\beta^{(m)},\eta^{(m)});k\right).
\end{equation*}
\State Update $\eta^{(m)}$ as
\begin{equation*}
\eta^{(m+1)}\in\h\left(\eta^{(m)}-\frac{1}{\bar{L}_{\eta}}\nabla_{\eta}f(\beta^{(m+1)},\eta^{(m)});n-h\right).
\end{equation*}
\end{algsubstates}
\State Fix the active sets
\begin{equation*}
J=\left\{j\in[p]:\beta^{(m)}_j\neq0\right\}\quad\text{and}\quad I=\left\{i\in[n]:\eta^{(m)}_i=0\right\},
\end{equation*}
and solve the low-dimensional convex problem
\begin{equation*}
\min_{\beta,\eta}\quad f(\beta,\eta)\quad\st\quad\beta_j=0\,\forall\,j\notin J,\,\eta_i=0\,\forall\,i\in I.
\end{equation*}
\end{algorithmic}
\end{algorithm}
Algorithm \ref{alg:pbgd} first performs cyclic projected gradient descent updates until a convergence tolerance $\epsilon$ is satisfied. Upon convergence, the active set is fixed, and the coefficients are ``polished.'' The polishing step can be performed by a simple least squares fit restricted to the predictors $J$ and observations $I$. In the special case that $h=n$, the algorithm reduces to the projected gradient descent method in \textcite{Bertsimas2016a}.

We now establish some convergence properties of Algorithm \ref{alg:pbgd}, extending results and proof techniques from \textcite[Proposition 6 and Theorem 3.1]{Bertsimas2016a}. To this end, we begin by stating the following definitions for points of \eqref{eq:rss2} that are stationary and $\epsilon$-optimal stationary.
\begin{definition}
\label{def:stat}
The point $(\hat{\beta},\hat{\eta})$, with $\|\hat{\beta}\|_0\leq k$ and $\|\hat{\eta}\|_0\leq n-h$, is said to be a stationary point of the optimization problem \eqref{eq:rss2} if, for any $\bar{L}_\beta\geq L_\beta$ and any $\bar{L}_\eta\geq L_\eta$, it satisfies the fixed point equations
\begin{equation*}
\hat{\beta}\in\h\left(\hat{\beta}-\frac{1}{\bar{L}_\beta}\nabla_\beta f(\hat{\beta},\hat{\eta});k\right)\quad\text{and}\quad\hat{\eta}\in\h\left(\hat{\eta}-\frac{1}{\bar{L}_\eta}\nabla_\eta f(\hat{\beta},\hat{\eta});n-h\right).
\end{equation*}
Furthermore, $(\hat{\beta},\hat{\eta})$ is said to be an $\epsilon$-optimal stationary point if, for any $\epsilon>0$, it satisfies the inequalities
\begin{equation*}
\left\|\hat{\beta}-\h\left(\hat{\beta}-\frac{1}{\bar{L}_\beta}\nabla_\beta f(\hat{\beta},\hat{\eta});k\right)\right\|_2^2\leq\epsilon\quad\text{and}\quad\left\|\hat{\eta}-\h\left(\hat{\eta}-\frac{1}{\bar{L}_\eta}\nabla_\eta f(\hat{\beta},\hat{\eta});n-h\right)\right\|_2^2\leq\epsilon.
\end{equation*}
\end{definition}
With these definitions in mind, the convergence properties of Algorithm \ref{alg:pbgd} are given as follows.
\begin{proposition}
\label{prop:convg}
Let $\{(\beta^{(m)},\eta^{(m)})\}$ be a sequence generated by Algorithm \ref{alg:pbgd}. Then, for any $\bar{L}_\beta\geq L_\beta$ and any $\bar{L}_\eta\geq L_\eta$, the sequence $\{f(\beta^{(m)},\eta^{(m)})\}$ is decreasing, converges, and satisfies the inequality
\begin{equation}
\begin{split}
f(\beta^{(m)},\eta^{(m)})-&f(\beta^{(m+1)},\eta^{(m+1)}) \\
&\geq\frac{1}{2}(\bar{L}_\beta-L_\beta)\|\beta^{(m+1)}-\beta^{(m)}\|_2^2+\frac{1}{2}(\bar{L}_\eta-L_\eta)\|\eta^{(m+1)}-\eta^{(m)}\|_2^2.
\end{split}
\end{equation}
Furthermore, for any $\bar{L}_\beta>L_\beta$, any $\bar{L}_\eta>L_\eta$, and a stationary point $(\beta^\star,\eta^\star)$, the sequence $\{(\beta^{(m)},\eta^{(m)})\}$ satisfies the following inequality after running Algorithm \ref{alg:pbgd} for $M$ iterations:
\begin{equation*}
\underset{1\leq m\leq M}{\min}\left(\|\beta^{(m+1)}-\beta^{(m)}\|_2^2+\|\eta^{(m+1)}-\eta^{(m)}\|_2^2\right)\leq2\frac{f(\beta^{(1)},\eta^{(1)})-f(\beta^\star,\eta^\star)}{M\min(\bar{L}_\beta-L_\beta,\bar{L}_\eta-L_\eta)}.
\end{equation*}
\end{proposition}
Proposition \ref{prop:convg} establishes that Algorithm \ref{alg:pbgd} generates a convergent sequence of objective values for the robust subsets problem. In particular, it follows from the second inequality that the algorithm arrives at an $\epsilon$-optimal stationary point in $O(\frac{1}{\epsilon})$ iterations. We highlight that Proposition \ref{prop:convg} does not require any special conditions on the predictor matrix $X$, which may be contaminated.

\subsubsection{Neighborhood search}

Given an initial point $(\beta^{(0)},\eta^{(0)})$ satisfying the sparsity constraints on $\beta$ and $\eta$, Algorithm \ref{alg:pbgd} is guaranteed to converge. However, as with most nonconvex optimization problems, the choice of the initial point can impact the quality of the solution produced. In general, setting $\beta^{(0)}=0$ and $\eta^{(0)}=0$ does not result in satisfactory solutions. To this end, we apply a neighborhood search method that largely alleviates this issue. Such methods recently proved useful in \textcite{Mazumder2020} for the $\ell_1$ and $\ell_2$ regularized best subsets problems. For reasons to be explained, the neighborhood search method (as a byproduct of its design) also produces solutions to the robust subsets problem \eqref{eq:rss2} for an entire grid of values of the parameters $k$ and $h$. This set of fitted models produced by the method is useful in practice because the best predictive $(k,h)$ is typically unknown and needs to be chosen from a set of parameters, say, $K\times H$ with $K=\{k_1,\ldots,k_q\}$ and $H=\{h_1,\ldots,h_r\}$. For instance, given data with $n=100$ and $p=20$ that might contain up to 25\% contamination, it is natural to consider $K=\{0,\ldots,20\}$ and $H=\{75,80,\ldots,100\}$.

The algorithm is conceptually simple but slightly cumbersome to write down. To assist in this effort, $\beta(k_i,h_j)$ and $\eta(k_i,h_j)$ are taken to denote variables in the robust subsets problem \eqref{eq:rss2} with $k=k_i$ and $h=h_j$, and $\hat{\beta}(k_i,h_j)$ and $\hat{\eta}(k_i,h_j)$ as the corresponding solutions produced by Algorithm \ref{alg:pbgd}. We assume that $k_1<k_2<\cdots<k_q$ and $h_1<h_2<\cdots<h_r$. Algorithm \ref{alg:ns} presents the neighborhood search method.
\begin{algorithm}[H]
\caption{Neighborhood search}
\label{alg:ns}
\hspace*{\algorithmicindent} \textbf{Input:} $K=\{k_1,\ldots,k_q\}$, $H=\{h_1,\ldots,h_r\}$, and $\epsilon>0$.
\begin{algorithmic}[1]
\State For all $(i,j)\in[q]\times[r]$, run Algorithm \ref{alg:pbgd} initialized with
\begin{equation*}
\beta^{(0)}(k_i,h_j)=0\quad\text{and}\quad\eta^{(0)}(k_i,h_j)=0.
\end{equation*}
\State Repeat the following step for all $(i,j)\in[q]\times[r]$: \label{alg:nsupdate}
\begin{algsubstates}
\State Take the neighborhood of $(i,j)$ as
\begin{equation*}
\mathcal{N}(i,j)=\left\{a\in[q],b\in[r]:|i-a|+|j-b|=1\right\}.
\end{equation*}
For all $(a,b)\in\mathcal{N}(i,j)$, run Algorithm \ref{alg:pbgd} initialized with
\begin{equation*}
\beta^{(0)}(k_i,h_j)=\h\left(\hat{\beta}(k_a,h_b);k_i\right)\quad\text{and}\quad\eta^{(0)}(k_i,h_j)=\h\left(\hat{\eta}(k_a,h_b);n-h_j\right).
\end{equation*}
If the best solution obtained from the neighborhood initializations improves on the incumbent solution, update $\hat{\beta}(k_i,h_j)$ and $\hat{\eta}(k_i,h_j)$ with the best solution.
\end{algsubstates}
\State Repeat step \ref{alg:nsupdate} until successive changes in $\sum_{i=1}^q\sum_{j=1}^rf\left(\hat{\beta}(k_i,h_j),\hat{\eta}(k_i,h_j)\right)$ are $\epsilon$ small.
\end{algorithmic}
\end{algorithm}
Algorithm \ref{alg:ns} first computes an initial set of solutions corresponding to the parameter set $K\times H$ by running Algorithm \ref{alg:pbgd} initialized with zero vectors. In the second step, it progresses through $K\times H$, at each stage fixing $(k,h)$ at $(k_i,h_j)$ and initializing Algorithm \ref{alg:pbgd} with the solutions that neighbor $(k_i,h_j)$. The neighboring solutions are usually small perturbations to the support of the incumbent solution, which often leads to the discovery of new feasible solutions. The final step involves recursively iterating this update scheme until no further improvements can be made. It is our experience that approximately 10-20 rounds of updates are typically required to achieve convergence.

\subsection{Parameter choices}

\subsubsection{Big-M parameters}
\label{subsubsec:bigm}

To operationalize the mixed-integer program described in this section, it is necessary to choose suitable values of the Big-M parameters (e.g., $\mathcal{M}_\beta$). Large values of these parameters can lead to numerical issues and poor solver performance, and thus we wish to set them to values as small as reasonably possible. A simple approach is to take $\mathcal{M}_\beta=\tau\|\hat{\beta}\|_\infty$ for some $\tau\geq1$, where $\hat{\beta}$ is a solution obtained from the heuristics. The same estimation process can be applied for other Big-M parameters.

By estimating the Big-M parameters using the process described here, there exists a possibility of excluding the true optimal solution from the feasible set. It remains an open research question how to estimate provably correct parameters in the absence of any assumptions on $X$. As an alternative to Big-M constraints, our implementation also supports SOS constraints. SOS constraints never exclude the true optimal solution but are less numerically efficient.

\subsubsection{Sparsity and robustness parameters}

The choice of $k$ and $h$ plays a critical role. Choosing a value of $k$ that is too large leads the estimator to overfit to the data, and choosing a value of $k$ that is too small leads the estimator to underfit. Similarly, choosing a large $h$ makes the estimator susceptible to breakdown, and choosing a small $h$ leads to a loss of efficiency (because it might lead to discarding good data). Given that we are interested in building predictive models, cross-validation can address these issues. However, because the data might be contaminated, standard cross-validation metrics such as mean square prediction error are inappropriate. A suitable alternative is trimmed (mean square) prediction error, which trims a portion of the largest square errors. Letting $e_{(1)},\ldots,e_{(n)}$ denote prediction errors ordered by absolute value, the trimmed prediction error can be written as
\begin{equation*}
\text{Trimmed prediction error}:=\frac{\sum_{i=1}^{[(1-\alpha)n]}e_{(i)}^2}{[(1-\alpha)n]},
\end{equation*}
where the trimming parameter is typically taken conservatively as $\alpha\in\{0.25,0.5\}$. This metric can be computed over a grid of candidate values of $(k,h)$ using the cross-validation errors.

There are alternatives to cross-validating $h$. \textcite{Alfons2013} applied a two-step procedure where $h$ is initially set to a fixed proportion of $n$ and then revised upward based on the magnitude of the residuals. However, this approach involves additional parametric assumptions and is numerically expensive like cross-validation. An avenue of future work is to investigate efficient procedures, e.g., estimating $h$ and $\beta$ jointly as can be done with Huber loss for $\sigma$ (the scale parameter) and $\beta$ \parencite{Lambert-Lacroix2011}.

\section{Breakdown point}
\label{sec:breakdown}

This section discusses the robustness of robust subset selection in terms of its finite-sample breakdown point. The notion of a finite-sample breakdown point originated with \textcite{Donoho1983} and has since become a standard measure for robustness in the casewise contamination framework. Roughly speaking, the breakdown point of an estimator is the minimum fraction of contaminated observations required to corrupt the estimator arbitrarily badly. A formal definition is given as follows.
\begin{definition}
Let $(X,Y)$ be an uncontaminated sample of size $n$, and let $(\tilde{X},\tilde{Y})$ be the same sample with $1\leq m\leq n$ observations replaced arbitrarily. Let $\Theta(X,Y)$ be an estimator given the sample $(X,Y)$. Then the finite-sample breakdown point of $\Theta(X,Y)$ is defined as
\begin{equation*}
b(\Theta;X,Y):=\underset{1\leq m\leq n}{\min}\left\{\frac{m}{n}:\underset{(\tilde{X},\tilde{Y})}{\sup}~\|\Theta(X,Y)-\Theta(\tilde{X},\tilde{Y})\|_2=\infty\right\}.
\end{equation*}
\end{definition}
We take $\Theta(X,Y)$ to be the objective value of the robust subsets estimator. Thus, with the above definition in mind, the main result of this section is written as follows.
\begin{theorem}
\label{the:breakdown}
Let $(X,Y)$ be a sample of size $n$, and let $\Theta(X,Y)$ be the optimal objective value to the robust subset selection problem \eqref{eq:rss} with $h\leq n$. Then $\Theta(X,Y)$ has the finite-sample breakdown point
\begin{equation*}
b(\Theta;X,Y)=\frac{n-h+1}{n}.
\end{equation*}
\end{theorem}
The proof follows steps similar to those used in the proof of the breakdown point in \textcite{Bertsimas2014} for the objective value of the least quantile of squares estimator.

It follows from Theorem \ref{the:breakdown} that robust subset selection can withstand up to $n-h$ contaminated observations. Moreover, because fixing $h=n$ yields the best subsets estimator, it follows that the breakdown point of best subset selection is $1/n$, meaning it is not robust to any level of contamination in the data. These results are consistent with experimental evidence provided in the following section.

We close this section with three remarks pertaining to the parameter $h$:
\begin{itemize}
\item The performance of the solver is related to the choice of $h$. The most conservative choice $h=[0.5n]$ is also the most computationally cumbersome. Therefore, it is desirable in terms of computation to choose a value of $h$ that is as large as reasonably possible.
\item If the analyst is comfortable that the data contain no more than 25\% contamination, taking $h=[0.75n]$ is generally accepted as a good compromise between efficiency and robustness \parencite{Rousseeuw2006}.
\item Using cross-validation as outlined in the previous section can alleviate the need to manually set $h$. Nonetheless, the trimming parameter $\alpha$ in the cross-validation metric remains to be chosen, but 25\% again seems a judicious choice.
\end{itemize}

\section{Experiments}
\label{sec:experiments}

We perform a series of numerical experiments on synthetic data to evaluate the performance of our estimator and algorithms in a variety of scenarios. The experiments of Section \ref{subsec:estimators} compare robust subsets with existing estimators, while those of Section \ref{subsec:algorithms} compare different algorithmic approaches for its computation.

In support of these exercises, the methods described in Section \ref{sec:computational} were implemented in the \texttt{R} package \texttt{robustsubsets}. Our package calls \texttt{Gurobi} as the mixed-integer solver and implements Algorithms \ref{alg:pbgd} and \ref{alg:ns} in \texttt{C++}. The package first runs neighborhood search over a specified parameter grid $K\times H$ and then runs mixed-integer optimization using the warm starts and variable bounds from neighborhood search. The data are standardized to have zero median and unit (normalized) median absolute deviation in advance of fitting the model. To obtain the best subsets estimator, the set $H$ is taken as $\{n\}$, and the data are standardized to have zero mean and unit standard deviation. The final model fits are returned on the original scale of the data.

All experiments are carried out using \texttt{R} 4.1.0 and \texttt{Gurobi} 9.1.2.

\subsection{Comparisons of estimators}
\label{subsec:estimators}

\subsubsection{Setup}

We study the linear model
\begin{equation*}
Y=X\beta^0+\varepsilon,\quad\varepsilon\sim N(0,I\sigma^2),
\end{equation*}
with the entries of the coefficient vector $\beta^0$ drawn randomly from $\{-1,0,1\}$ and the number of nonzero coefficients $p_0:=\|\beta^0\|_0\in\{5,10\}$. The rows of the predictor matrix $X$ are sampled iid as $x_i\sim N(0,\Sigma)$, where $\Sigma$ has row $i$ and column $j$ constructed as $0.35^{|i-j|}$ for all $i,j\in[p]$. The noise variance $\sigma^2$ is chosen to yield the desired signal-to-noise ratio (SNR), where
\begin{equation*}
\text{SNR}:=\frac{\var(x^T\beta^0)}{\sigma^2}=\frac{(\beta^0)^T\Sigma\beta^0}{\sigma^{2}}.
\end{equation*}
We take $\text{SNR}\in\{1,4\}$ when $p_0=5$ and $\text{SNR}\in\{4,9\}$ when $p_0=10$, with $\text{SNR}=1$ corresponding to 50\% proportion of variance explained (PVE), where
\begin{equation*}
\text{PVE}:=\frac{\var(x^T\beta^0)}{\var(y)}=\frac{\text{SNR}}{\text{SNR}+1}.
\end{equation*}
We study a low-dimensional setup in which $n=500$ and $p=100$, and a high-dimensional setup in which $n=100$ and $p=500$.

We consider four contamination settings:
\begin{enumerate}
\item No contamination - The response and predictors are both uncontaminated.
\item Contamination of $Y$ - The response is contaminated by sampling the noise as a mixture of normal distributions: $\varepsilon_i\sim (1-\delta) N(0,\sigma^2)+\delta N(10\sigma,\sigma^2)$, $i\in[n]$.
\item Contamination of $X$ - The rows of the predictor matrix are first sampled as $N(0,\Sigma)$ and the response is generated. Each row of $X$ is subsequently contaminated with probability $\delta$ by randomly selecting $0.1 p$ predictors and replacing their values with independent draws from a $N(10,1)$ distribution.
\item Contamination of $Y$ and $X$ - The response and predictors are both contaminated as described above.
\end{enumerate}
The contamination probability $\delta=0.1$. The expected number of contaminated observations in the sample $(X,Y)$ is thus $0.1n$ under settings two and three and $0.19n$ under setting four. Settings one, two, and four are similar to those studied in \textcite{Chang2018}. Setting three is added to evaluate the effect of contamination in $X$ alone.

To benchmark robust subsets, three contemporary sparse and robust estimators are also evaluated: the Lasso with Tukey's bisquare loss (MM Lasso) via \texttt{pense} 1.2.9 \parencite{Smucler2017,CohenFreue2019}, the Lasso with LTS loss (sparse LTS) via \texttt{robustHD} 0.6.1 \parencite{Alfons2013}, and the Lasso with Huber loss (Huber Lasso) via \texttt{hqreg} 1.4 \parencite{Rosset2007,Yi2017}. The availability of high-quality implementations of these estimators underscores their relevance to practitioners. We also evaluate the vanilla (least squares) Lasso as implemented in \texttt{glmnet} 4.1-1.

For robust subsets, the tuning parameters $k$ and $h$ are swept over the grid $K\times H$ with $K=\{0,\ldots,20\}$ and $H=\{[0.75n],[0.80n],\ldots,n\}$. For best subsets, $k$ is swept over the same $K$. For the Lasso estimators, we sweep the tuning parameter $\lambda$ over 50 values linearly spaced on the log scale, with the maximum $\lambda$ set according to the default of each package.

To measure expected \emph{out-of-sample} prediction loss, we study the relative prediction error:
\begin{equation*}
\text{Relative prediction error}:=\frac{\e[(y-\hat{\mu}-x^T\hat{\beta})^2]}{\sigma^2}=\frac{(\beta^0-\hat{\beta})^T\Sigma(\beta^0-\hat{\beta})+\hat{\mu}^2+\sigma^2}{\sigma^2},
\end{equation*}
where $\hat{\mu}$ is an estimate of the intercept (the true intercept is zero), and $\hat{\beta}$ is an estimate of $\beta^0$. The best attainable relative prediction error is 1 and the null relative prediction error is $\text{SNR}+1$. We also study the sparsity of the fitted model:
\begin{equation*}
\text{Model sparsity}:=\|\hat{\beta}\|_0,
\end{equation*}
and, to measure support recovery, the F1 score:
\begin{equation*}
\text{F1 score}:=\frac{2}{\text{Recall}^{-1}+\text{Precision}^{-1}},
\end{equation*}
which is the harmonic average of recall (the true positive rate) and precision (the positive predictive value). The best attainable F1 score is 1, indicating that the support of $\hat{\beta}$ exactly matches that of $\beta^0$. \textcite{Hastie2020} considered these three metrics (relative prediction error, model sparsity, and F1 score) in their comparisons of best subsets and the Lasso. The metrics are all evaluated with respect to tuning parameters chosen via 10-fold cross-validation.\footnote{The parameter $h$ in sparse LTS is not treated as a tuning parameter in \texttt{robustHD}, and is instead fixed at 75\% of the sample size. After the model is initially fit, \texttt{robustHD} applies a reweighting step to improve efficiency. The reader is referred to \textcite{Alfons2013} for details.} The cross-validation metrics are (mean square) prediction error for the nonrobust estimators and trimmed prediction error with 25\% trimming for the robust estimators. For best subsets and robust subsets, only the heuristics are used during cross-validation to maintain reasonable runtime. Mixed-integer optimization is run on the $k$ and $h$ yielding the lowest cross-validation error with variable bounds estimated by the method of Section \ref{subsubsec:bigm} using $\tau=1.5$.

\subsubsection{Results}

We conduct 30 simulations for each set of simulation parameters and then aggregate the results. The simulations are performed in parallel, each running on a single core of an AMD Ryzen Threadripper 3970x. In the interest of space, we confine the results and discussion here to sparsity level $p_0=5$ and relegate those for $p_0=10$ to the appendix.

\begin{figure}[!ht]
\centering
\input{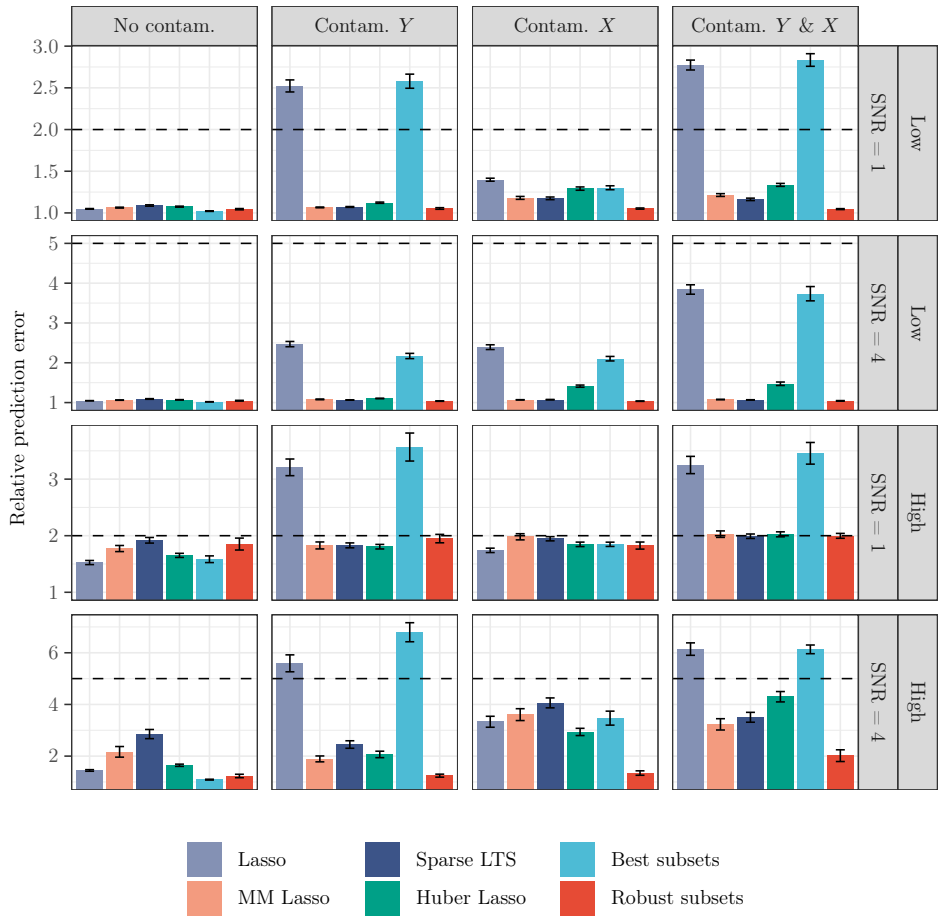}
\caption{Relative prediction error estimated over 30 simulations with $p_0=5$. The vertical bars represent averages, and the error bars denote (one) standard errors. The dashed horizontal lines indicate the relative prediction error from the null model.}
\label{fig:staterror5}
\end{figure}

Figures \ref{fig:staterror5}, \ref{fig:statsparsity5}, and \ref{fig:statf1score5} report the relative prediction error, model sparsity, and F1 score, respectively. The vertical bars represent averages, and the error bars denote (one) standard errors. The dashed horizontal lines in Figure \ref{fig:staterror5} indicate the relative prediction error from the null model, and those in Figure \ref{fig:statsparsity5} indicate the sparsity of the true model.
\begin{figure}[!ht]
\centering
\input{Figures/Statistical-Figure-Sparsity-5.tex}
\caption{Model sparsity estimated over 30 simulations with $p_0=5$. The vertical bars represent averages, and the error bars denote (one) standard errors. The dashed horizontal lines indicate the true model sparsity.}
\label{fig:statsparsity5}
\end{figure}
\begin{figure}[!ht]
\centering
\input{Figures/Statistical-Figure-F1Score-5.tex}
\caption{F1 score estimated over 30 simulations with $p_0=5$. The vertical bars represent averages, and the error bars denote (one) standard errors.}
\label{fig:statf1score5}
\end{figure}

In the uncontaminated settings, best subsets exhibits excellent prediction accuracy and support recovery. The Lasso, while also showing excellent prediction accuracy, has inferior support recovery compared with best subsets. The F1 scores of the Lasso are hindered by the large number of irrelevant predictors it picks up. When contamination is introduced, both best subsets and the Lasso display significant performance degradation across the board. Only in the Low-4 configuration (low-dimensional setup with $\text{SNR}=4$), where there are relatively few predictors and the signal is strong, do they consistently outperform the null model in terms of prediction accuracy.

In the contaminated settings, robust subsets ameliorates the degradation in performance that occurs in best subsets. Robust subsets behaves in a manner similar to that of best subsets as if it were applied to a reduced set of ``good'' observations. Similarly, the MM Lasso, sparse LTS, and the Huber Lasso largely retain the operational characteristics of the Lasso, though the latter struggles with contamination in $X$. In terms of prediction accuracy, robust subsets produces fitted models that are competitive with and often superior to those from the robust Lasso estimators. The main success story is the High-4 configuration, in which robust subsets improves markedly on its competitors across all contamination settings.

Robust subsets inherits the good support recovery qualities of best subsets. In almost all cases, robust subsets enjoys the highest F1 score among the robust estimators. Upon closer inspection, the robust Lasso estimators perform slightly better in detecting true positives. However, they pay a steep price in the number of false positives they produce, leading to lower F1 scores overall. They also produce relatively dense models. For instance, when $X$ is contaminated, the MM Lasso selects up to 60 predictors, 12 times the true number of nonzeros. On the other hand, robust subsets consistently delivers models that more closely reflect the true sparsity level.

Generally, the robust estimators struggle more with contamination in the predictors than in the response. In fact, in the High-1 configuration, when $X$ is contaminated, all robust estimators fail to predict better than their nonrobust counterparts and offer little to no improvement on the null model. This result suggests that building good predictive models may be an unachievable goal when the signal is weak and the number of contaminated predictors is large. However, when the signal is strong, as in the High-4 configuration, robust subsets is able to offer improvement. In fact, robust subsets performs about as well with regard to support recovery as it does in the low-dimensional setups.

\subsection{Comparisons of algorithms}
\label{subsec:algorithms}

\subsubsection{Setup}

The previous experiments compare robust subsets with existing estimators---the experiments below provide further insight into the algorithms underlying robust subsets. The simulation design remains as before. We focus on the high-dimensional setup, fixing $n=100$ and taking $p\in\{500,1000\}$. The number of nonzeros $p_0=5$ and $\text{SNR}=4$. We consider contamination setting two where $Y$ is contaminated and setting four where $X$ is also contaminated. The total proportion of contaminated observations in both settings is fixed at 10\%, with contamination evenly split between $Y$ and $X$ in setting four.

To isolate the individual contributions of each algorithm in our framework, we consider several different computational approaches: neighborhood search ($K$ and $H$ specified as before) without mixed-integer optimization, mixed-integer optimization without warm starts or variable bounds, and mixed-integer optimization with warm starts and variable bounds from neighborhood search. These approaches are respectively labeled ``heuristics,'' ``MIO,'' and ``MIO+heuristics.'' The variable bounds for the last approach are again estimated using the method of Section \ref{subsubsec:bigm} with $\tau\in\{1,1.5\}$. The MIO approach uses SOS constraints in place of Big-M constraints since variable bounds are unavailable. The solver is run on an AMD Ryzen Threadripper 3970x with a 30 minute time limit.

To measure the quality of the solution produced, we study the number of true positive predictors selected and the relative objective gap:
\begin{equation*}
\text{Relative objective gap}:=\frac{\hat{f}-f^\star}{f^\star},
\end{equation*}
where $\hat{f}$ is the attained objective value, and $f^\star$ is the best objective value among all approaches considered. To measure progress towards proving optimality, we study the relative optimality gap:
\begin{equation*}
\text{Relative optimality gap}:=\frac{\hat{f}-\hat{f}_L}{\hat{f}},
\end{equation*}
where $\hat{f}_L$ is the lower bound on the optimal objective value delivered by the solver.\footnote{This definition is used in \texttt{Gurobi}.} A relative optimality gap of zero indicates that the computed solution is provably optimal. Additionally, we measure whether the solver terminated within the time limit (i.e., had an optimality gap of zero) and the runtime. These metrics are evaluated at $k=5$ and $h=90$.

\subsubsection{Results}

Table \ref{tab:comp500} reports results from 30 simulations for $p=500$. Results for $p=1000$ are detailed in the appendix. Averages or proportions are reported with standard errors in parentheses.
\begin{table}[!ht]
\centering
\footnotesize
\begin{tabular}{llllll}
\toprule
 & True pos. & Obj. gap (\%) & Opt. gap (\%) & Term. (\%) & Time (mins.) \\ 
\midrule
\multicolumn{6}{l}{Contamination of $Y$} \\ 
\midrule
Heuristics & 5.0 (0.0) & 0.0 (0.0) & - & - &  0.5 (0.0) \\ 
MIO & 5.0 (0.0) & 0.0 (0.0) & 100.0 (0.0) &   0.0 (0.0) & 30.1 (0.0) \\ 
MIO+heuristics ($\tau=1$) & 5.0 (0.0) & 0.0 (0.0) &   0.0 (0.0) & 100.0 (0.0) &  0.9 (0.1) \\ 
MIO+heuristics ($\tau=1.5$) & 5.0 (0.0) & 0.0 (0.0) &  25.5 (5.4) &  53.3 (9.1) & 19.1 (2.1) \\ 
\midrule
\multicolumn{6}{l}{Contamination of $Y$ and $X$} \\ 
\midrule
Heuristics & 4.9 (0.1) & 1.9 (1.6) & - & - &  2.5 (0.1) \\ 
MIO & 5.0 (0.0) & 0.0 (0.0) & 100.0 (0.0) &  0.0 (0.0) & 30.1 (0.0) \\ 
MIO+heuristics ($\tau=1$) & 5.0 (0.0) & 0.0 (0.0) &  30.5 (5.4) & 33.3 (8.6) & 26.3 (1.8) \\ 
MIO+heuristics ($\tau=1.5$) & 5.0 (0.0) & 0.0 (0.0) &  96.3 (1.2) &  0.0 (0.0) & 32.6 (0.1) \\ 
 \bottomrule
\end{tabular}

\caption{True positive selections, relative objective gap, relative optimality gap, termination rate, and runtime estimated over 30 simulations with $n=100$, $p=500$, $p_0=5$, and $\text{SNR}=4$. Averages or proportions are reported next to (one) standard errors in parentheses.}
\label{tab:comp500}
\end{table}

When only $Y$ is contaminated, all algorithmic approaches are equally effective at delivering high-quality solutions as measured by the number of true positive selections and the objective gap. There is, however, significant disparity in terms of the optimality gap. Without warm starts or variable bounds, the solver is never able to improve the lower bound in the 30 minute time limit. Yet, when guided by this information from the heuristics, the solver proves optimality in 100\% of the simulation instances for $\tau=1$, typically within a minute. These same figures are 53\% and 19 minutes for $\tau=1.5$, confirming that small values of $\tau$ play an important role in determining the speed at which the optimality gap is closed.

When $Y$ and $X$ are contaminated, the heuristic solutions are slightly lower-quality. The three mixed-integer approaches continue to generate excellent solutions. For $\tau=1$, the solver attains a zero optimality gap in a third of the simulation instances. The optimality gap is much looser for $\tau=1.5$. More generous time limits (roughly several hours to tens of hours) are required to close the optimality gap in most cases. The heuristics now take longer to converge than when $Y$ is contaminated, 2 minutes more on average. Likewise, the solver takes longer to progress the lower bound regardless of the value for $\tau$. These longer runtimes and weaker optimality gaps can be attributed to poorer conditioning of the gram matrix $X^TX$.

The speed at which the solver is able to close the optimality gap can also be impacted by the degree to which observations are outlying. This impact is a consequence of the fact that the Big-M parameter $\mathcal{M}_\eta$ (the outlier bound) must be of the same order of magnitude as the outliers. Another contributing factor is that the condition number of the gram matrix is affected by the size of the outliers in $X$. Overall, smaller outliers mean tighter optimality gaps and smaller runtimes. Larger outliers have the opposite effect.

Additional results in the appendix show that the proposed methods scale to $p=1000$ in reasonable time. It is difficult, however, to close the optimality gap in the 30 minute time limit. Nevertheless, when warm-started, the solver usually finds the optimal solution in the first few minutes, while the remaining time is spent proving optimality. Thus, if the problem instance appears intractable, it can be sufficient to run the solver for a short time to obtain a high-quality solution. Scaling to instances with $p$ (or $n$) in the tens or hundreds of thousands while proving optimality is a direction for future research.

Besides the type of contamination and number of predictors, which we vary in these experiments, several other factors determine the benefit of running the solver after the heuristics. Though it is difficult to quantify exact gains in general, our experience is that mixed-integer optimization is most useful when the number of contaminated observations, number of nonzero coefficients, or levels of correlation among predictors are high. The SNR does not appear to be an important factor.

\section{Archaeological glass vessels dataset}
\label{sec:archaeological}

This section illustrates an application of robust subsets using the archaeological glass vessels dataset introduced in \textcite{Janssens1998} and \textcite{Lemberge2000} and studied vis-\`{a}-vis sparsity and robustness in \textcite{Smucler2017} and \textcite{Amato2021}. The dataset was obtained from the supplemental material of \textcite{Christidis2020}. It consists of observations on 180 glass vessels from the 16th and 17th centuries that were uncovered in archaeological excavations in Antwerp, Belgium. To understand the origins of the vessels and the trade connections between producers, the chemical compositions of the vessels were studied. To determine these compositions, electron-probe X-ray microanalysis (EPXMA) was used to produce data containing EPXMA intensities for 1920 different frequencies on each of the glass vessels. The data were subsequently processed to yield the concentrations of chemical compounds present in the vessels. However, the processing step is time-consuming and challenging to automate, and thus interest exists in regression methods that can directly predict the concentrations using the EPXMA data. The majority of estimators do not perform well at this task because the data are high-dimensional and include multiple observations that constitute outliers \parencite{Serneels2005,Maronna2011}.

\begin{figure}[!ht]
\centering
\input{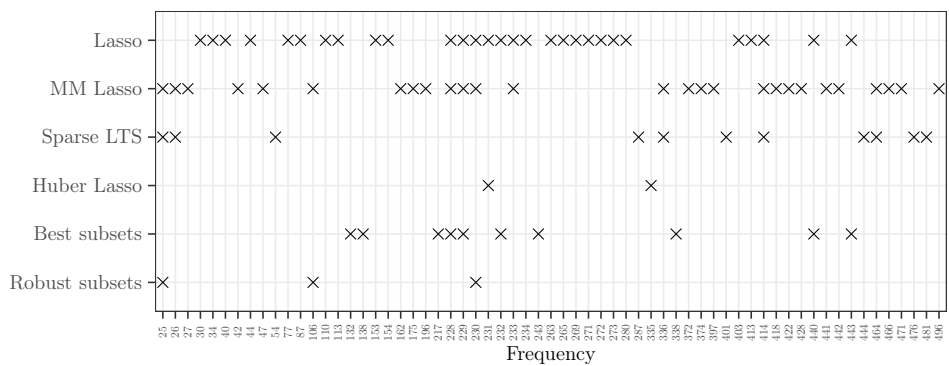}
\caption{Selected frequencies (predictors) for the archaeological glass vessels dataset. The marks identify the frequencies with nonzero coefficients in the fitted models.}
\label{fig:appselected}
\end{figure}

Following \textcite{Smucler2017}, the response is taken as the concentration of the chemical compound PbO and the predictors as the frequencies 15 through 500. Frequencies outside this range have little variation and are almost null. The resulting sample has $n=180$ and $p=486$. Each estimator is applied to this full sample. The Lasso produces a fitted model containing 29 frequencies. The MM Lasso selects 27 frequencies, sparse LTS selects 11, and the Huber Lasso selects 2. Best subsets selects 10 frequencies, while robust subsets produces a model with 3. To glean insight into these fitted models, we present Figure \ref{fig:appselected}, which indicates the nonzero coefficients in each model.
It is apparent from this figure that the robust subsets and best subsets models do not overlap at all. The story is largely similar for the Lasso, with only 6 of its 29 predictors shared with its robust adaptions. While the Huber Lasso delivers a highly sparse model, its frequencies are not shared with the other robust estimators, possibly because it does not try to resist contamination in $X$. On the other hand, every frequency picked up by robust subsets is also picked up by the MM Lasso.

\begin{figure}[!ht]
\centering
\input{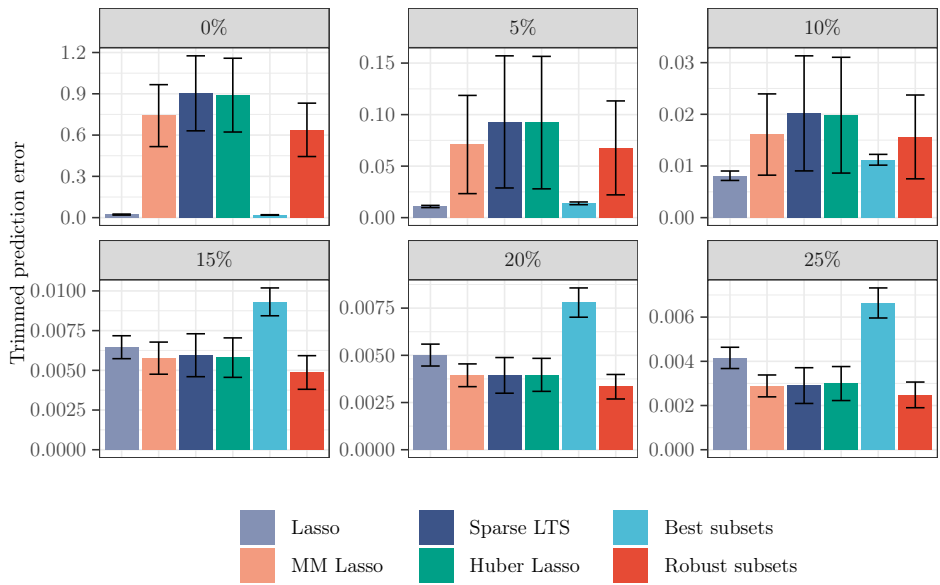}
\caption{Trimmed prediction error, expressed as a function of the trimming level, estimated via 10-fold cross-validation for the archaeological glass vessels dataset. The vertical bars represent averages, and the error bars denote (one) standard errors.}
\label{fig:apperror}
\end{figure}

To evaluate the prediction accuracy of the competing estimators, we use 10-fold cross-validation and record the trimmed prediction error. Figure \ref{fig:apperror} reports this metric across several levels of trimming to accommodate various severities of contamination. The vertical bars represent averages, and the error bars denote (one) standard errors.
For low levels of trimming, the nonrobust estimators dominate in terms of prediction error. This result is not particularly surprising because any outliers will inflate the prediction error. The transition point at which all estimators fare similarly occurs between the 10\% and 15\% trimming levels. At higher levels of trimming, the robust estimators outperform their nonrobust counterparts. In particular, robust subsets improves substantially on best subsets and yields the smallest prediction error on average among the competing estimators.

\section{Concluding remarks}
\label{sec:concluding}

Best subset selection is a classic tool for sparse regression, and recent developments in mathematical optimization have paved the way for exciting new research into this estimator. Inspired by these developments, this paper proposes robust subset selection, an adaption of best subset selection that it is resistant to casewise contaminated data. The combinatorial problem that defines robust subsets is shown to be amenable to mixed-integer optimization, a technology that continues to display tremendous improvements. To speed up runtime, heuristic methods that complement the mixed-integer optimization approach are developed. Central to the heuristics is a projected block-coordinate gradient descent method, for which we derive convergence properties. As a statistical guarantee, the objective value of the robust subsets estimator is shown to resist a specifiable level of contamination in finite samples. Numerical experiments on synthetic and real data yield findings consistent with this result. The ability of best subsets to recover the true support and produce good predictions is observed to deteriorate significantly if the data are contaminated. In contrast, robust subsets resists contamination by excluding from the model fit the subset of observations that induce the most substantial losses. Compared with robust adaptions of continuous shrinkage estimators, robust subsets does well to closely recover the underlying sparsity pattern. This property makes robust subsets a promising tool for applications in which the fitted models themselves are of interest and not just their prediction accuracy alone.

Our implementation \texttt{robustsubsets} is available as an \texttt{R} package at
\begin{center}
\url{https://github.com/ryan-thompson/robustsubsets}.
\end{center}

\section*{Acknowledgments}

The author thanks Catherine Forbes and Farshid Vahid for their encouragement and suggestions, and the anonymous referees for their constructive comments. This research was supported by an Australian Government Research Training Program (RTP) Scholarship.

\printbibliography

\begin{appendices}

\section{Computational methods}

\subsection{Improved problem formulations}

It is possible to reformulate the robust subsets program \eqref{eq:rssmio} to yield improved computational performance. Consider the following mixed-integer program:
\begin{equation}
\label{eq:rssmio2}
\begin{split}
\underset{\xi,\beta,\eta,s,z}{\min}\quad&\frac{1}{2}\sum_{i=1}^n(y_i-\xi_i)^2 \\
\st\quad&\xi=X\beta+\eta \\
&-\mathcal{M}_\xi\leq\xi_i\leq\mathcal{M}_\xi,\qquad i\in[n] \\
&s_j\in\{0,1\},\qquad j\in[p] \\
&-\mathcal{M}_\beta\leq\beta_j\leq\mathcal{M}_\beta,\qquad j\in[p] \\
&-s_j\mathcal{M}_{\beta}\leq\beta_j\leq s_j\mathcal{M}_{\beta},\qquad j\in[p] \\
&\sum_{j=1}^ps_j\leq k \\
&z_i\in\{0,1\},\qquad i\in[n] \\
&-\mathcal{M}_{\eta}\leq\eta_i\leq \mathcal{M}_{\eta},\qquad i\in[n] \\
&-z_i\mathcal{M}_{\eta}\leq\eta_i\leq z_i\mathcal{M}_{\eta},\qquad i\in[n] \\
&\sum_{i=1}^nz_i\leq n-h,
\end{split}
\end{equation}
where $\xi\in\mathbb{R}^n$ is an additional auxiliary variable. Observe that the objective functions in \eqref{eq:rssmio} and \eqref{eq:rssmio2} differ only in the number of variables involved. Specifically, the objective in \eqref{eq:rssmio2} is a function of $p+n$ variables, whereas the objective in \eqref{eq:rssmio2} is a function of $n$ variables only. In terms of computational performance, it is our experience that mixed-integer solvers respond better to \eqref{eq:rssmio2} because it has fewer quadratic terms. The above formulation also adds additional structure to the mixed-integer program by bounding the $\ell_\infty$-norms of $\beta$ and $\eta$ by $\mathcal{M}_\beta$ and $\mathcal{M}_\eta$ via the bound constraints $-\mathcal{M}_\beta\leq\beta_j\leq\mathcal{M}_\beta$ and $-\mathcal{M}_{\eta}\leq\eta_i\leq\mathcal{M}_{\eta}$. The Big-M constraints imply these bounds. There are a number of other implied bounds that can also be added to the program. The reader is referred to \textcite{Bertsimas2016a} for further details.

\subsection{Proof of Proposition \ref{prop:convg}}

The proof proceeds along the lines of that for Proposition 6 and Theorem 3.1 in \textcite{Bertsimas2016a}. 

We begin by proving the first part of Proposition \ref{prop:convg}. Let $\hat{\beta}$ denote an update to any $k$-sparse $\beta$:
\begin{equation*}
\hat{\beta}\in\h\left(\beta-\frac{1}{\bar{L}_\beta}\nabla_\beta f(\beta,\eta);k\right),
\end{equation*}
and take $\bar{L}_\beta\geq L_\beta$, an upper bound to the partial Lipschitz constant. Then, from Lemma \eqref{lemma:descent}, we have the following series of inequalities:
\begin{equation*}
\begin{split}
f(\beta,\eta)&=Q(\beta,\beta) \\
&\geq\underset{\|\tilde{\beta}\|_0\leq k}{\inf}~Q(\tilde{\beta},\beta) \\
&=\underset{\|\tilde{\beta}\|_0\leq k}{\inf}\left(f(\beta,\eta)+\nabla_\beta f(\beta,\eta)^T(\tilde{\beta}-\beta)+\frac{1}{2}\bar{L}_\beta\|\tilde{\beta}-\beta\|_2^2\right) \\
&=\underset{\|\tilde{\beta}\|_0\leq k}{\inf}\left(f(\beta,\eta)-\frac{1}{2\bar{L}_\beta}\|\nabla_\beta f(\beta,\eta)\|_2^2+\frac{1}{2}\bar{L}_\beta\left\|\tilde{\beta}-\left(\beta-\frac{1}{\bar{L}_\beta}\nabla_\beta f(\beta,\eta)\right)\right\|_2^2\right) \\
&=f(\beta,\eta)-\frac{1}{2\bar{L}_\beta}\|\nabla_\beta f(\beta,\eta)\|_2^2+\frac{1}{2}\bar{L}_\beta\left\|\hat{\beta}-\left(\beta-\frac{1}{\bar{L}_\beta}\nabla_\beta f(\beta,\eta)\right)\right\|_2^2 \\
&=f(\beta,\eta)+\nabla_\beta f(\beta,\eta)^T(\hat{\beta}-\beta)+\frac{1}{2}\bar{L}_\beta\|\hat{\beta}-\beta\|_2^2 \\
&=f(\beta,\eta)+\nabla_\beta f(\beta,\eta)^T(\hat{\beta}-\beta)+\frac{1}{2}L_\beta\|\hat{\beta}-\beta\|_2^2+\frac{1}{2}(\bar{L}_\beta-L_\beta)\|\hat{\beta}-\beta\|_2^2 \\
&\geq f(\hat{\beta},\eta)+\frac{1}{2}(\bar{L}_\beta-L_\beta)\|\hat{\beta}-\beta\|_2^2.
\end{split}
\end{equation*}
Taking $\beta=\beta^{(m)}$, $\hat{\beta}=\beta^{(m+1)}$, and $\eta=\eta^{(m)}$, it follows that
\begin{equation}
\label{eq:convg1}
f(\beta^{(m)},\eta^{(m)})-f(\beta^{(m+1)},\eta^{(m)})\geq\frac{1}{2}(\bar{L}_\beta-L_\beta)\|\beta^{(m+1)}-\beta^{(m)}\|_2^2.
\end{equation}
Similarly, letting $\hat{\eta}$ denote an update to any $(n-h)$-sparse $\eta$:
\begin{equation*}
\hat{\eta}\in\h\left(\eta-\frac{1}{\bar{L}_\eta}\nabla_\eta f(\beta,\eta);n-h\right),
\end{equation*}
and applying Lemma \eqref{lemma:descent} with $\eta=\eta^{(m)}$, $\hat{\eta}=\eta^{(m+1)}$, and $\beta=\beta^{(m+1)}$, we obtain
\begin{equation}
\label{eq:convg2}
f(\beta^{(m+1)},\eta^{(m)})-f(\beta^{(m+1)},\eta^{(m+1)})\geq\frac{1}{2}(\bar{L}_\eta-L_\eta)\|\eta^{(m+1)}-\eta^{(m)}\|_2^2.
\end{equation}
Adding together \eqref{eq:convg1} and \eqref{eq:convg2} yields
\begin{equation}
\label{eq:convg3}
\begin{split}
f(\beta^{(m)},\eta^{(m)})-&f(\beta^{(m+1)},\eta^{(m+1)}) \\
&\geq\frac{1}{2}(\bar{L}_\beta-L_\beta)\|\beta^{(m+1)}-\beta^{(m)}\|_2^2+\frac{1}{2}(\bar{L}_\eta-L_\eta)\|\eta^{(m+1)}-\eta^{(m)}\|_2^2.
\end{split}
\end{equation}
Hence, the sequence $\{f(\beta^{(m)},\eta^{(m)})\}$ is decreasing, and because $f(\beta,\eta)$ is bounded below by zero, it follows from the monotone convergence theorem that the sequence converges.

For the second part of Proposition \ref{prop:convg}, we take the sum of \eqref{eq:convg3} over $1\leq m\leq M$ to obtain
\begin{equation}
\label{eq:convg4}
\begin{split}
\sum_{m=1}^M&\left(f(\beta^{(m)},\eta^{(m)})-f(\beta^{(m+1)},\eta^{(m+1)})\right) \\
&\geq\frac{1}{2}\sum_{m=1}^M\left((\bar{L}_\beta-L_\beta)\|\beta^{(m+1)}-\beta^{(m)}\|_2^2+(\bar{L}_\eta-L_\eta)\|\eta^{(m+1)}-\eta^{(m)}\|_2^2\right).
\end{split}
\end{equation}
The inequality \eqref{eq:convg4} implies that
\begin{equation*}
\begin{split}
f(&\beta^{(1)},\eta^{(1)})-f(\beta^{(M+1)},\eta^{(M+1)}) \\
&\geq\frac{M}{2}\underset{1\leq m\leq M}{\min}\left((\bar{L}_\beta-L_\beta)\|\beta^{(m+1)}-\beta^{(m)}\|_2^2+(\bar{L}_\eta-L_\eta)\|\eta^{(m+1)}-\eta^{(m)}\|_2^2\right) \\
&\geq\frac{M}{2}\min(\bar{L}_\beta-L_\beta,\bar{L}_\eta-L_\eta)\underset{1\leq m\leq M}{\min}\left(\|\beta^{(m+1)}-\beta^{(m)}\|_2^2+\|\eta^{(m+1)}-\eta^{(m)}\|_2^2\right).
\end{split}
\end{equation*}
Because $\{f(\beta^{(m)},\eta^{(m)})\}$ is decreasing and converges to $f(\beta^\star,\eta^\star)$, it follows that
\begin{equation*}
\begin{split}
\underset{1\leq m\leq M}{\min}\left(\|\beta^{(m+1)}-\beta^{(m)}\|_2^2+\|\eta^{(m+1)}-\eta^{(m)}\|_2^2\right)&\leq2\frac{f(\beta^{(1)},\eta^{(1)})-f(\beta^{(M+1)},\eta^{(M+1)})}{M\min(\bar{L}_\beta-L_\beta,\bar{L}_\eta-L_\eta)} \\
&\leq2\frac{f(\beta^{(1)},\eta^{(1)})-f(\beta^\star,\eta^\star)}{M\min(\bar{L}_\beta-L_\beta,\bar{L}_\eta-L_\eta)},
\end{split}
\end{equation*}
with the final inequality that which we set out to obtain.

\section{Breakdown point}

\subsection{Proof of Theorem \ref{the:breakdown}}

The proof below follows steps similar to those used in the proof of the breakdown point in \textcite{Bertsimas2014} for the objective value of the least quantile of squares estimator. We use the following standard result in the proof.
\begin{lemma}
\label{lemma:optval}
Let $\Theta(X,Y)$ be the optimal objective value to the robust subset selection problem \eqref{eq:rss}. Then $\Theta(X,Y)$ satisfies the equality
\begin{equation*}
\Theta(X,Y)=\underset{I\in\mathcal{I}}{\min}~\underset{\beta\in\mathrm{B}}{\min}~\frac{1}{2}\sum_{i\in I}(y_i-x_i^T\beta)^2,
\end{equation*}
where
\begin{equation*}
\mathcal{I}=\{I\subseteq[n]:|I|\geq h\}\quad\text{and}\quad\mathrm{B}=\{\beta\in\mathbb{R}^p:\|\beta\|_0\leq k\}.
\end{equation*}
\end{lemma}
We proceed by completing the proof of Theorem \ref{the:breakdown} in two parts, showing that the inequalities $b(\Theta;X,Y)>(n-h)/n $ and $b(\Theta;X,Y)\leq(n-h+1)/n$ both hold. The former inequality is proven first. Suppose that exactly $m=n-h$ observations of the original sample $(X,Y)$ are arbitrarily contaminated, and denote this new contaminated sample $(\tilde{X},\tilde{Y})$. Let $I^0$ contain only the indices of the uncontaminated observations. Because $I^0\in\mathcal{I}$, it follows from Lemma \ref{lemma:optval} that
\begin{equation}
\label{eq:breakdown1}
\Theta(\tilde{X},\tilde{Y})=\underset{I\in\mathcal{I}}{\min}~\underset{\beta\in\mathrm{B}}{\min}~\frac{1}{2}\sum_{i\in I}(\tilde{y}_i-\tilde{x}_i^T\beta)^2\leq\underset{\beta\in\mathrm{B}}{\min}~\frac{1}{2}\sum_{i\in I^0}(\tilde{y}_i-\tilde{x}_i^T\beta)^2.
\end{equation}
The right-hand side of \eqref{eq:breakdown1} does not depend on any contaminated observations and is finite. Thus, the breakdown point is strictly larger than $(n-h)/n$. Suppose that one additional observation is arbitrarily contaminated such that $m=n-h+1$. Therefore, every $I\in\mathcal{I}$ includes a contaminated observation, say the observation indexed by $c$. Let $I^\star$ and $\beta^\star$ denote an optimal solution to the robust subsets problem \eqref{eq:rss}. Then the optimal objective value is lower bounded as
\begin{equation}
\label{eq:breakdown2}
\Theta(\tilde{X},\tilde{Y})=\frac{1}{2}\sum_{i\in I^\star}(\tilde{y}_i-\tilde{x}_i^T\beta^\star)^2\geq\frac{1}{2}(\tilde{y}_c-\tilde{x}_c^T\beta^\star)^2.
\end{equation}
The right-hand side of \eqref{eq:breakdown2} can be made arbitrarily large because $\tilde{y}_c$ can be made arbitrarily large. Thus, the breakdown point is less than or equal to $(n-h+1)/n$. We conclude that $b(\Theta;X,Y)=(n-h+1)/n$.

\section{Experiments}

\subsection{Comparisons of estimators}

\begin{figure}[!ht]
\centering
\input{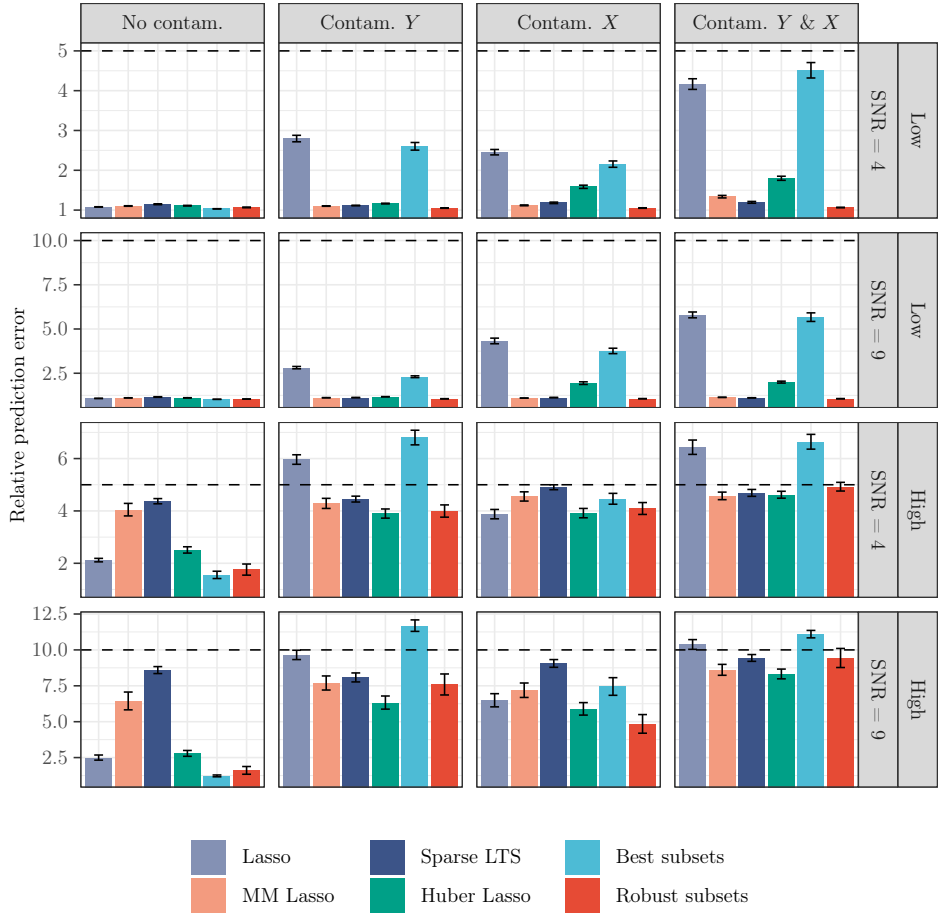}
\caption{Relative prediction error estimated over 30 simulations with $p_0=10$. The vertical bars represent averages, and the error bars denote (one) standard errors. The dashed horizontal lines indicate the relative prediction error from the null model.}
\label{fig:staterror10}
\end{figure}

The results of Section \ref{subsec:estimators} for $p_0=10$ nonzero coefficients are reported in Figures \ref{fig:staterror10}, \ref{fig:statsparsity10}, and \ref{fig:statf1score10}. The findings in the low-dimensional setup are broadly consistent with those for $p_0=5$, while the high-dimensional setup proves more formidable. Robust subsets maintains superior support recovery when either $Y$ or $X$ are contaminated and performs within statistical precision when both are contaminated. However, when both are contaminated, none of the estimators offer more than marginal improvement in prediction over the null model, even when $\text{SNR}=9$.
\begin{figure}[!ht]
\centering
\input{Figures/Statistical-Figure-Sparsity-10.tex}
\caption{Model sparsity estimated over 30 simulations with $p_0=10$. The vertical bars represent averages, and the error bars denote (one) standard errors. The dashed horizontal lines indicate the true model sparsity.}
\label{fig:statsparsity10}
\end{figure}
\begin{figure}[!ht]
\centering
\input{Figures/Statistical-Figure-F1Score-10.tex}
\caption{F1 score estimated over 30 simulations with $p_0=10$. The vertical bars represent averages, and the error bars denote (one) standard errors.}
\label{fig:statf1score10}
\end{figure}

\subsection{Comparisons of algorithms}

The results of Section \ref{subsec:algorithms} for $p=1000$ predictors are reported in Table \ref{tab:comp1000}. When $Y$ and $X$ are contaminated, and in the absence of warm start information, the solution from the solver is typically low-quality. The combination of heuristics and mixed-integer optimization continues to produce the best outcomes across the board. In both contamination settings, setting $\tau=1$ produces optimality gaps that are far superior to those from setting $\tau=1.5$. As with $p=500$, using the smaller value of $\tau=1$ does not harm the number of true positive selections or the objective gap.
\begin{table}[!ht]
\centering
\footnotesize
\begin{tabular}{llllll}
\toprule
 & True pos. & Obj. gap (\%) & Opt. gap (\%) & Term. (\%) & Time (mins.) \\ 
\midrule
\multicolumn{6}{l}{Contamination of $Y$} \\ 
\midrule
Heuristics & 5.0 (0.0) & 0.2 (0.2) & - & - &  1.3 (0.1) \\ 
MIO & 5.0 (0.0) & 0.0 (0.0) & 100.0 (0.0) &  0.0 (0.0) & 30.0 (0.0) \\ 
MIO+heuristics ($\tau=1$) & 5.0 (0.0) & 0.0 (0.0) &   0.7 (0.7) & 96.7 (3.3) &  3.7 (1.0) \\ 
MIO+heuristics ($\tau=1.5$) & 5.0 (0.0) & 0.0 (0.0) &  71.1 (5.4) & 10.0 (5.5) & 30.5 (0.6) \\ 
\midrule
\multicolumn{6}{l}{Contamination of $Y$ and $X$} \\ 
\midrule
Heuristics & 4.9 (0.1) &  1.8 (1.3) & - & - &  7.8 (0.4) \\ 
MIO & 3.8 (0.3) & 30.5 (8.7) & 100.0 (0.0) & 0.0 (0.0) & 30.0 (0.0) \\ 
MIO+heuristics ($\tau=1$) & 5.0 (0.0) &  0.0 (0.0) &  68.8 (3.8) & 3.3 (3.3) & 37.3 (0.7) \\ 
MIO+heuristics ($\tau=1.5$) & 5.0 (0.0) &  0.0 (0.0) & 100.0 (0.0) & 0.0 (0.0) & 37.9 (0.4) \\ 
 \bottomrule
\end{tabular}

\caption{True positive selections, relative objective gap, relative optimality gap, termination rate, and runtime estimated over 30 simulations with $n=100$, $p=1000$, $p_0=5$, and $\text{SNR}=4$. Averages or proportions are reported next to (one) standard errors in parentheses.}
\label{tab:comp1000}
\end{table}

\end{appendices}

\end{document}